\documentclass[aps,showpacs]{revtex4-2}
\usepackage{graphicx,color}
\def\no{\noindent}
\def\bc{\begin{center}}
\def\ec{\end{center}}

\def\beq{\begin{equation}}
\def\eeq{\end{equation}}

\def\br{{\bf r}}

\def\bR{{\bf R}}

\def\bk{{\bf k}}

\begin{document}

\title{Localization by particle-hole symmetry breaking:\\
a loop expansion
}

\author{K. Ziegler}
\address{Institut f\"ur Physik, Universit\"at Augsburg\\
D-86135 Augsburg, Germany}
\date{\today}

\begin{abstract}
Localization by a broken particle-hole symmetry in a random system of non-interacting quantum particles is studied
on a $d$--dimensional lattice.  Our approach is based on a chiral symmetry argument and the corresponding 
invariant measure, where the latter is described by a Grassmann functional integral.
Within a loop expansion we find for small loops diffusion in the case of particle-hole symmetry. 
Breaking the particle-hole symmetry results in the creation of random dimers, which suppress
diffusion and lead to localization on the scale $\sqrt{D/|\mu|}$, where $D$ is the effective diffusion coefficient
at particle-hole symmetry and $\mu$ is the parameter related to particle-hole symmetry breaking.  
\end{abstract}

\maketitle

\section{Introduction}

There is strong evidence that disordered systems with particle-hole (PH) symmetry can avoid Anderson localization in 
any spatial dimension. Such a behavior was observed for one-dimensional systems
at the band center of a one-dimensional tight-binding model with random hopping 
some time ago~\cite{PhysRev.92.1331,PhysRevB.13.4597,PhysRevB.18.569}. 
There is also numerical evidence for extended states at the band center in disordered two-dimensional lattice 
models~\cite{PhysRevB.26.1838,PhysRevB.49.3190}. A renewed interest in this problem appeared with the discovery
of two-dimensional Dirac-like materials, such as 
graphene~\cite{no.ge.mo.ji.ka.gr.du.fi.05,RevModPhys.81.109,ab.ap.be.zi.ch.10}. 
Graphene is a semimetal with a very robust conductivity at the PH-symmetric Dirac point. On the other hand, 
after breaking the PH symmetry by doping, its conductivity changes substantially; it is either enhanced for weak 
disorder or reduced for strong 
disorder~\cite{Chen2008,doi:10.1126/science.1167130,PhysRevLett.102.236805,PhysRevLett.103.056404}.
The effect of disorder on topological materials, based on Dirac-like models, has also been the subject of some recent 
theoretical research~\cite{PhysRevB.95.155122,PhysRevLett.123.046801,PIXLEY2021168455}.
Despite of a substantial effort, there was 
no conclusive confirmation of localization away from the PH-symmetric Dirac point, neither from the 
theory side \cite{RevModPhys.83.407,PhysRevB.73.125411} nor from the 
experiment~\cite{Chen2008,doi:10.1126/science.1167130,PhysRevLett.102.236805,PhysRevLett.103.056404}.

While in the PH-symmetric case diffusion was identified as the dominant behavior of non-interacting Dirac particles
in a random environment, the breaking of the PH symmetry was accompanied with the creation of random 
dimers~\cite{zi_13}. A preliminary work, based on a perturbative renormalization group analysis, did not 
reveal localization, though~\cite{si.zi.12}. 
In the following we will analyze the competition of diffusing quantum particles and randomly distributed dimers on
a $d$--dimensional lattice without employing a perturbation theory.

Starting from a PH-symmetric random Hamiltonian $H$, a symmetry-breaking term is defined by a uniform shift $\mu$
of the PH-symmetric Hamiltonian. The resulting Hamiltonian is still invariant under a chiral transformation. 
This invariance is relevant for the analysis of spontaneous chiral-symmetry breaking and the corresponding long-scale
behavior of the model, such that we reduce our description to the corresponding invariant measure (IM). 
Then the IM, written in terms of a Grassmann functional integral~\cite{negele2019quantum}, is
represented by a loop expansion, which consists of graphs with 4-vertices. Only the smallest loops are
taken into account, which is known as the nonlinear sigma model approximation. This is used as the starting
point for the analysis of the spatial correlations. As mentioned above, there is diffusion in the presence of PH symmetry
with a diffusion coefficient $D$, which depends on the scattering rate $\eta$ and the average Hamiltonian
$\langle H\rangle$. After breaking the PH symmetry with $\mu\ne 0$, we create small loops with two sites on the
lattice which represent repulsive lattice dimers. They act as obstacles for the diffusion and suppress the latter on large
scales, which leads to an exponential decay of the particle correlation. Since the density of the dimers is proportional
to $|\mu|$, this localization effect increases with this parameter. An estimation of the decay length gives
$\sqrt{D/|\mu|}$.

\section{Model: Invariant measure}

We briefly recapitulate the main ideas which were developed in the previous work on the loop expansion
and the IM for non-interacting particles in a random environment~\cite{PhysRevB.79.195424,zieg.15}.
For this purpose a random Hamiltonian matrix $H$ on a lattice $\Lambda$ is considered. It is assumed that 
this Hamiltonian has
an internal spinor structure; its matrix elements are of the form $H_{\br j,\br' j'}$, where $\br,\br'\in \Lambda$ and
$j,j'=1,2,...,N$ is a spinor index, where the latter can also be a band index of a multiband Hamiltonian. 
We further assume that there is an $N\times N$ unitary matrix $U$ that (i) acts only on the spinor or band index and 
(ii) for which the Hamiltonian matrix obeys the relation
\beq
U H^T U^\dagger=-H
\ .
\label{id0}
\eeq
This relation implies a particle-hole transformation in the following sense: Since $H$ is Hermitian, we have
$H^T=H^*$ and the relation (\ref{id0}) implies for the eigenstate $\Psi_E$ of $H$ with the real eigenvalue $E$
\[
H^*U^\dagger\Psi_E=-U^\dagger H\Psi_E=-EU^\dagger\Psi_E
\ .
\]
Complex conjugation of this equation yields
\[
H(U^\dagger\Psi_E)^*=-E(U^\dagger\Psi_E)^*
\ ,
\]
such that $\Psi_{-E}=(U^\dagger\Psi_E)^*$ is eigenstate of $H$ with eigenvalue $-E$. Thus, there is a PH symmetry
for $E=0$. The PH symmetry 
of $H$ is broken by $\mu$ for $H_\pm:=H\pm\mu\sigma_0$, where $\sigma_0$ is the $N\times N$ unit matrix.
In the following we will analyze the effect of a shift of $E=0$ by $\mu$, using the nonlinear sigma model approach.

Following Refs. \cite{si.zi.12,zi_13}, we extend $H$ to the random Hamiltonian matrix
\beq
\label{8by8a}
{\hat H}=\pmatrix{
H_+ & 0 & 0 & 0\cr
0 & H_- & 0 & 0\cr
0 & 0 & H_+^T & 0 \cr
0 & 0 & 0 & H_-^T \cr
}
\equiv\pmatrix{
{\bar H}  & 0\cr
0 & {\bar H}^T\cr
}
\ .
\eeq
At the PH symmetric point ($\mu=0$) it satisfies the relation ${\hat H}{\hat S}_0=-{\hat S}_0{\hat H}$ for
\beq
{\hat S}_0=\pmatrix{
0 & 0 & \varphi_{3}U & \varphi_{1}U \cr
0 & 0 & \varphi_{2}U & \varphi_{4}U \cr
\varphi'_{3}U^\dagger & \varphi'_{2}U^\dagger & 0 & 0 \cr
\varphi'_{1}U^\dagger & \varphi'_{4}U^\dagger & 0 & 0 \cr
}
\label{unbroken_s}
\eeq
with some general parameters $\varphi_j$. 
For a broken PH symmetry ($\mu\ne 0$) it satisfies ${\hat H}{\hat S}=-{\hat S}{\hat H}$ for
\beq
{\hat S}=\pmatrix{
0 & 0 & 0 & \varphi_{1}U \cr
0 & 0 & \varphi_{2}U & 0\cr
0 & \varphi'_{2}U^\dagger & 0 & 0 \cr
\varphi'_{1}U^\dagger & 0 & 0 & 0 \cr
}
\ .
\label{broken_s}
\eeq
The PH-symmetric case was studied previously, such that we can focus subsequently on the broken PH symmetry.
Then the relation ${\hat H}{\hat S}=-{\hat S}{\hat H}$
implies for ${\hat H}$ the chiral symmetry $e^{\hat S}{\hat H}e^{\hat S}={\hat H}$ of the extended Hamiltonian 
defined in Eq. (\ref{8by8a}).
The chiral symmetry reveals some interesting properties, which will be discussed next. 

Some general remarks on the IM: The goal is to calculate average quantities, such as the average Green's function or 
the average product of two Green's functions, with respect to the random matrix elements of the Hamiltonian ${\hat H}$
\cite{PhysRevB.79.195424}. This should be seen as an alternative to studies, where the distribution of the random 
Hamiltonian and its spectrum is 
considered~\cite{Liaw2013,https://doi.org/10.1002/ctpp.201700111,PhysRevResearch.2.043375}.
Average quantities are sufficient to discuss many physically motivated questions, such as transport \cite{zi_13,zieg.15}.
Although simpler than the analysis of the random distribution, the averaging with respect to the random Hamiltonian
is a tedious task for a large lattice $\Lambda$. A common and often successful approximation 
to this problem is to perform a saddle-point integration (also known as the method of steepest 
descent)~\cite{garfken67:math}.
Then another problem occurs when there is no unique saddle-point
solution but a manifold of saddle points due to some symmetry of the Hamiltonian. A typical example is the Hamiltonian
(\ref{8by8a}) with the chiral symmetry. Since all the saddle points of the manifold are equally important, we must 
integrate over all of them. The resulting saddle-point integral leads to the IM. This will be discussed in detail for the specific
example of ${\hat H}$ subsequently. The saddle-point integral can be performed in this case and leads to a sum that is
characterized by loops with increasing size. There is no problem with convergence on a finite lattice, since the loops are
strictly repulsive and the size of the loops is restricted by the size of the lattice.
The loop expansion was previously developed for the PH-symmetric case in Refs.~\cite{PhysRevB.79.195424,zieg.15}
and will be adopted to the PH-symmetry broken case in the following. 

First, we construct the IM that is associated with the chiral symmetry.
For the matrix ${\hat S}$ and the graded determinant (cf. Eq. (\ref{detg0})) we get
\[
detg(e^{2{\hat S}})=\exp[Trg( \log e^{2{\hat S}})]=\exp[2Trg({\hat S})]=1
\ ,
\]
since the graded trace vanishes: $Trg({\hat S})=0$.
Together with the chiral symmetry of ${\hat H}$ this implies immediately
\beq
\label{im0000}
detg({\hat H}+i\eta)=detg({\hat H}+i\eta)detg(e^{2{\hat S}})
=detg(e^{\hat S}{\hat H}e^{\hat S}+i\eta e^{2{\hat S}})
\]
\[
=detg({\hat H}+i\eta e^{2{\hat S}})
\ ,
\eeq
where $detg({\hat H}+i\eta)=1$, according to the definition of the graded determinant. 
Thus, $detg({\hat H}+i\eta e^{2{\hat S}})$ is invariant under the chiral transformation.

From relation (\ref{im0000}) we can construct the IM through substituting the general parameters 
$\varphi_j$ by a spatial Grassmann field $\varphi_{\br j}$. As explained in App. \ref{sect:no_sb0}, this leads to the 
lattice version of the IM with
\beq
\label{IM0001}
J_j=\det({\bf 1}+\varphi'_j\varphi_j -\varphi'_jh\varphi_j h^\dagger)^{-1}
=\det({\bf 1}+\varphi_j\varphi'_j -h\varphi_j h^\dagger\varphi'_j)
\ \ \ (j=1,2),
\eeq
where the relation is derived in (\ref{inversion}), with
\beq
\label{eff_green}
h={\bf 1}+2i\eta{\bar G}_0
\ ,\ \ \
{\bar G}_0=({\bar H}_0+i\epsilon -i\eta)^{-1}
\ .
\eeq
$J_j$ is an invariant for a {\it global} chiral transformation, provided that $h$ is unitary. This is the case for $\epsilon=0$:
$hh^\dagger={\bf 1}+O(\epsilon)$ \cite{zieg.15}.
It is important to realize that the random Hamiltonian ${\hat H}$ has been replaced by its average 
${\bar H}_0:=\langle {\bar H}\rangle$ in the IM.
In Refs. \cite{PhysRevB.79.195424,zieg.15} this IM was associated with the correlator $K_{\bR\bR'}$ between the lattice 
sites $\bR$ and $\bR'$ through the relation
\beq
\label{corr00}
K_{\bR\bR'}
=\frac{1}{\cal N}\int_\Lambda J_j\varphi_{\bR j}\varphi'_{\bR' j} 
\ , \ \ \ 
{\cal N}=\int_\Lambda J_1=\int_\Lambda J_2 
\eeq
of a two-component Grassmann field $\varphi_{\br j}$ ($\br\in\Lambda$).  $\int_\Lambda$ is the functional integral 
with respect to the Grassmann field on the lattice $\Lambda$.
This correlator indicates localization on the localization length $\xi$ when it decays exponentially on the scale $\xi$.
It is identical for both components of the Grassmann field $j=1,2$, such that we can drop this index subsequently.
Another important feature of the Green's function $h$ is that the relation (\ref{id0}) implies for the PH-symmetric case 
$\mu=0$ the relation
\beq
\label{ph_symm1}
Uh^\dagger U^\dagger=h
\ ,
\eeq
which does not hold for $\mu\ne 0$.

The Grassmann field can be expressed by its Fourier components as
\beq
\varphi_\br=\sum_\bk e^{-i\bk\cdot\br}{\tilde\varphi}_\bk
={\tilde\varphi}_0+\sum_{\bk\ne 0} e^{-i\bk\cdot\br}{\tilde\varphi}_\bk\equiv {\tilde\varphi}_0 +\tau_{\br}
\ ,
\eeq
where the zero mode ${\tilde\varphi}_0$ does not depend on $\br$. Then the normalization becomes
\[
{\cal N}=\int_\Lambda\det\{{\bf 1}+[{\tilde\varphi}_0 +\tau-h({\tilde\varphi}_0 +\tau)h^\dagger]
({\tilde\varphi}_0' +\tau')\}
\]
and with $hh^\dagger={\bf 1}-\epsilon\bar{\Gamma}+O(\epsilon^2)$ we get
\[
{\tilde\varphi}_0-h{\tilde\varphi}_0h^\dagger
={\tilde\varphi}_0({\bf 1} -hh^\dagger)
=\epsilon\bar{\Gamma} {\tilde\varphi}_0+O(\epsilon^2)
\ ,
\]
since the space-independent ${\tilde\varphi}_0$ commutes with $h$.
$\epsilon$ can be absorbed into the Grassmann integration by rescaling ${\tilde\varphi}_0\to {\bar\tau}/\epsilon$.
This implies for the normalization
\beq
\label{normalization00}
{\cal N}=\epsilon\int_\Lambda\det[{\bf 1}+(\bar{\Gamma}{\bar\tau}+\tau-h\tau h^\dagger)
({\tilde\varphi}_0' +\tau')]
\ ,
\eeq
where we have neglected terms of order $\epsilon$ inside the integrand.
For the unnormalized correlator this rescaling argument for ${\tilde\varphi}_0$ provides a constant term 
${\tilde\varphi}_0{\tilde\varphi}'_0$ from the external Grassmann variable $\varphi_\bR\varphi'_{\bR'}$:
\beq
\int_\Lambda J\varphi_{\bR}\varphi'_{\bR'} =\int_\Lambda J({\tilde\varphi}_0+\tau_{\bR})({\tilde\varphi}'_0+\tau'_{\bR'})
\]
\[
=\int_\Lambda J{\bar\tau}({\tilde\varphi}'_0+\tau'_{\bR'})+
\epsilon\int_\Lambda J\tau_{\bR}({\tilde\varphi}'_0
+\tau'_{\bR'})
\ ,
\eeq
where only the first term on the right-hand side vanishes with $\epsilon\to0$. This implies for the normalized
correlator that the second term diverges for $\epsilon\to0$, which is a consequence of the broken 
translational invariance due to the factor $\tau_\bR$.

\section{Loop expansion}

The IM of Eq. (\ref{IM0001}) can be rewritten as
\beq
J=\exp\{Tr([WW^\dagger]_d\varphi\varphi')\}\det(
{\bf 1}-\varphi X^\dagger\varphi'-X\varphi\varphi'-W\varphi W^\dagger\varphi')
\eeq
when we define $h=[h]_d+W$, $X=W[h]_d^\dagger$, where $[h]_d$ and $[WW^\dagger]_d$ are the spatial 
diagonal parts of $h$ and $WW^\dagger$, respectively. Here we implicitly assume the limit $\epsilon\to0$. 
Using the determinant identity $\det(A)=\exp[Tr(\log A)]$ this yields for the IM after expanding the logarithm
\beq
\label{loop_expansion00}
\log J=
Tr([WW^\dagger]_d\varphi\varphi')
-\sum_{l\ge 1}\frac{1}{l}Tr[(\varphi X^\dagger\varphi'+X\varphi\varphi'+W\varphi W^\dagger\varphi')^l]
\ .
\eeq
This sum terminates on a finite lattice $\Lambda$ for $l=|\Lambda|$ due to the Grassmann field. The trace term with 
the power $l$ represents a sum of loops of length $l$ on the lattice, such that the sum can be considered as a loop 
expansion of the IM~\cite{PhysRevB.79.195424}. 
Integration with respect to the Grassmann field yields graphs with 4-vertices from the term 
$W\varphi W^\dagger\varphi'$ because the non-zero Grassmann integral requires at each site $\br$ the product 
$\varphi_\br\varphi'_\br$. The 4-vertex graphs reflect the equivalence of the expression (\ref{loop_expansion00})
with the random phase representation of the IM \cite{zieg.15}.

\subsection{Nonlinear sigma model}
\label{sect:nonlinear_sigma}

As a special case of the loop expansion (\ref{loop_expansion00}) only the smallest loops are considered, namely
only loops that contain at most two hopping matrices, either $X$ or $W$.
This approximation is known as the nonlinear sigma model \cite{1960NCim...16..705G} and becomes in the present case
\[
\log J_{NLSM}=
\]
\[
Tr([WW^\dagger]_d\varphi\varphi')
-Tr(W\varphi W^\dagger\varphi')
-Tr(\varphi X^\dagger\varphi'+X\varphi\varphi')
\]
\[
-Tr[(\varphi X^\dagger\varphi'+X\varphi\varphi')(\varphi X^\dagger\varphi'+X\varphi\varphi')]
\ ,
\]
where the third term vanishes due to the trace of an off-diagonal matrix $X$, such that the IM reduces to
\beq
\label{nlsm0}
\log J_{NLSM}=Tr([WW^\dagger]_d\varphi\varphi')
-Tr(W\varphi W^\dagger\varphi')
\]
\[
+Tr(X^\dagger\varphi\varphi' X^\dagger\varphi \varphi'-X\varphi\varphi'X\varphi\varphi')
\ .
\eeq
The first two terms represent a diffusion propagator
\beq
Tr_N([WW^\dagger]_{\br\br})\delta_{\br\br'}-Tr_N(W_{\br\br'}W^\dagger_{\br'\br})
\]
\[
=\sum_{\br''}Tr_N(W_{\br\br''}W^\dagger_{\br''\br})\delta_{\br\br'}-Tr_N(W_{\br\br'}W^\dagger_{\br'\br})
\ ,
\eeq
where $Tr_N$ is the trace with respect to the spinor index.
The quartic term reads
\[
Tr(X^\dagger\varphi\varphi' X^\dagger\varphi \varphi'-X\varphi\varphi'X\varphi\varphi')
=\sum_{\br,\br'}\beta_{\br\br'}\varphi_{\br}\varphi'_{\br}\varphi_{\br'}\varphi'_{\br'}
\ .
\]
$\beta$ is an imaginary matrix due to 
\beq
\label{loc_matrix}
\beta_{\br\br'}:=Tr_N(X^\dagger_{\br\br'}X^\dagger_{\br'\br}-X_{\br\br'}X_{\br'\br})
\ .
\eeq
Relation (\ref{ph_symm1}) implies that $\beta$ vanishes in the presence of the PH symmetry 
(i.e., for $\mu=0$). This was also observed in Refs. \cite{si.zi.12,zi_13}.

Next, we represent the functional integral with a quadratic form in the Grassmann field, using a coupling of the latter 
to a real Gaussian field $Q_\br$. This can be achieved by exploiting the relation
\beq
m_\br {\bar\beta}_{\br\br'}m_{\br'}
-(Q_\br +V_{\br\br''}m_{\br''})V^{-1}_{\br\br'}(Q_{\br'} +V_{\br'\br'''}m_{\br'''})
=-Q_\br V^{-1}_{\br\br'}Q_{\br'}
-2m_\br Q_\br
\eeq
with the sum convention for paired indices and
with the correlation matrix $V=i\alpha{\bf 1}+{\bar\beta}$ and ${\bar\beta}=\beta/i|\mu|$. We note that 
$m_\br V_{\br\br'}m_{\br'}=m_\br {\bar\beta}_{\br\br'}m_{\br'}$ for Grassmann variables $m_\br$.
Then we introduce the Gaussian integral 
\beq
\label{Gauss_dist}
\langle ... \rangle_Q :=
\frac{(i|\mu|/\pi)^{|\Lambda|/2}}{\sqrt{\det V}}\int_Q\exp(-i|\mu|\sum_{\br,\br'}Q_\br V^{-1}_{\br\br'}Q_{\br'}) ...
\prod_{\br\in\Lambda}dQ_\br
\ ,
\eeq
where we have set the free positive parameter $\alpha$ such that $V$ is non-singular and its eigenvalues have positive
real parts. The Grassmann integration can be performed because the argument of the exponential function 
is a quadratic form of the Grassmann field:
\beq
\label{grassmann_int2}
K_{\bR\bR'}
=\frac{1}{{\cal N}}\Big\langle\int_\Lambda\exp\left\{
{\tilde\gamma}_0\sum_\br\varphi_{\br}\varphi'_{\br}
-\sum_{\br,\br'}\gamma_{\br\br'}\varphi_{\br}\varphi'_{\br'}
-2i|\mu|\sum_{\br\in\Lambda}Q_\br\varphi_{\br}\varphi'_{\br}\right\}\varphi_\bR\varphi'_{\bR'}
\Big\rangle_Q
\]\[
=\frac{1}{{\cal N}}\langle adj_{\bR\bR'}({\tilde\gamma}_0-\gamma-2i|\mu|Q)\rangle_Q
\ .
\eeq
with $\gamma_{\br\br'}=Tr_N(W_{\br\br'}W^\dagger_{\br'\br})$ and 
${\tilde\gamma}_0=\sum_{\br'}\gamma_{\br\br'}$. 
The adjugate matrix can be expressed by the determinant as
\beq
\label{adjugate_m}
adj_{\bR\bR'}({\tilde\gamma}_0-\gamma-2i|\mu|Q)
=\det({\tilde\gamma}_0-\gamma-2i|\mu|Q)({\tilde\gamma}_0-\gamma-2i|\mu| Q)^{-1}_{\bR\bR'}
\ ,
\eeq
provided that ${\tilde\gamma}_0-\gamma+2i|\mu|Q$ is not singular. The latter can always be  arranged by a deformation
of the $Q_\br$ path of integration in the complex plane. The deformation moves the poles of the inverse matrix 
away from the real axis, which results in an exponential decay with respect to $|\bR-\bR'|$.

\subsection{Estimation of the localization length}
\label{sect:local}

\begin{figure}
    \centering
    \includegraphics[width=3.5cm,height=2.5cm]{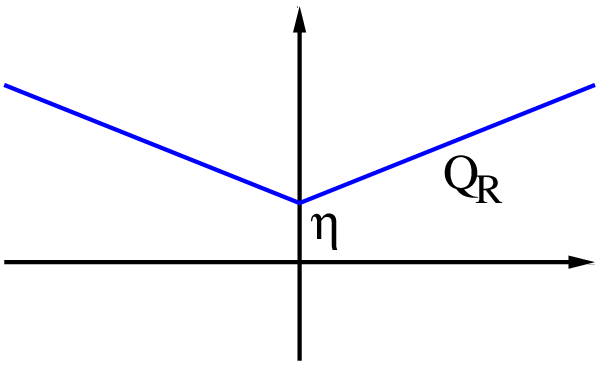}
    \caption{Contour $\Gamma$ of the $Q_\bR$ integration in the complex plane. The angle with respect to the
    abscissa is $\pm\zeta$ ($0<\zeta<\pi/4$).
    }    
\label{fig:1}
\end{figure}

To analyze the spatial behavior of the inverse matrix we use the plane wave eigenvector 
$\Phi_\bk=(\exp[i\bk\cdot\br])$ of the translational invariant matrix $\gamma$ with eigenvalue ${\tilde\gamma}_\bk$:
\beq
({\tilde\gamma}_0-\gamma-2i|\mu|Q)\Phi_\bk=({\tilde\gamma}_0-{\tilde\gamma}_\bk-2i|\mu|Q)\Phi_\bk
\eeq
with a diagonal matrix on the right-hand side. Since we have
\beq
\Phi_\bk=
({\tilde\gamma}_0-\gamma-2i|\mu|Q)^{-1}({\tilde\gamma}_0-\gamma-2i|\mu|Q)\Phi_\bk
\]
\[
=({\tilde\gamma}_0-\gamma-2i|\mu|Q)^{-1}({\tilde\gamma}_0-{\tilde\gamma}_\bk-2i|\mu|Q)\Phi_\bk
\ ,
\eeq
we get for $\Phi'_\bk:=({\tilde\gamma}_0-{\tilde\gamma}_\bk-2i|\mu|Q)\Phi_\bk$ the equation
\beq
\label{m-equ}
({\tilde\gamma}_0-\gamma-2i|\mu|Q)^{-1}\Phi'_\bk=\Phi_\bk
=({\tilde\gamma}_0-{\tilde\gamma}_\bk-2i|\mu|Q)^{-1}\Phi'_\bk
\ .
\eeq
Then we consider the basis $\{\Phi'_\bk\}$ and define the vector $\phi'_{\bR}:=\int_\bk \phi_{\bk,\bR}\Phi'_\bk$.
This gives with Eq. (\ref{m-equ})
\beq
({\tilde\gamma}_0-\gamma-2i|\mu|Q)^{-1}\phi'_{\bR}
=\int_\bk \phi_{\bk,\bR}({\tilde\gamma}_0-{\tilde\gamma}_\bk-2i|\mu|Q)^{-1}\Phi'_\bk
=\int_\bk\phi_{\bk,\bR}\Phi_\bk
\ .
\eeq
Using the special local vector $\phi_{\bR'}=(\delta_{\bR'\br})$, we get
\beq
\phi_{\bR'}\cdot({\tilde\gamma}_0-\gamma-2i|\mu|Q)^{-1}\phi'_{\bR}
=\int_\bk \phi_{\bk,\bR}(\phi_{\bR'}\cdot({\tilde\gamma}_0-{\tilde\gamma}_\bk-2i|\mu|Q)^{-1}\Phi'_\bk)
\]
\[
=\int_\bk\phi_{\bk,\bR}(\phi_{\bR'}\cdot\Phi_\bk)
=\int_\bk\phi_{\bk,\bR}e^{i\bk\cdot\bR'}
\ .
\eeq
Finally, we define the expansion coefficients as 
$\phi_{\bk,\bR}=e^{-i\bk\cdot\bR}/({\tilde\gamma}_0-{\tilde\gamma}_\bk-2i|\mu|Q_{\bR})$ and obtain
\beq
\phi_{\bR'}\cdot({\tilde\gamma}_0-\gamma-2i|\mu|Q)^{-1}\phi'_{\bR}
=\int_\bk\frac{e^{i\bk\cdot(\bR'-\bR)}}{{\tilde\gamma}_0-{\tilde\gamma}_\bk-2i|\mu|Q_{\bR}}
\ ,\ \ 
\]
\[
\phi'_{\bR}=\int_\bk \frac{e^{-i\bk\cdot\bR}}
{{\tilde\gamma}_0-{\tilde\gamma}_\bk-2i|\mu|Q_{\bR}}
({\tilde\gamma}_0-{\tilde\gamma}_\bk-2i|\mu|Q)\Phi_\bk
\ .
\eeq
The decay of this expression with respect to $|\bR-\bR'|$ characterizes the localization even before averaging,
i.e., for any realization of $Q_\bR$. 
To calculate the decay we  choose the contour $\Gamma$ of the $Q_\bR$ integration as
$Q_\bR=e^{i{\rm sgn}({\bar Q})\zeta}{\bar Q} +i\eta$ with real 
$-\infty<{\bar Q}<\infty$, $\eta>0$ and $0<\zeta<\pi/4$, as visualized in Fig. \ref{fig:1}. 
Moreover, we assume that ${\tilde\gamma}_0-{\tilde\gamma}_\bk=Dk^2$, which is the inverse diffusion propagator
with diffusion coefficient $D$.
Then the $k_1$ integration can be carried out for the decay along the 1--direction of the lattice as
\beq
\int_\bk\frac{e^{i\bk\cdot(\bR'-\bR)}}{{\tilde\gamma}_0-{\tilde\gamma}_\bk-2i|\mu|Q_\bR}
=\int_{\bk_\perp}\int_{-\infty}^\infty \frac{e^{ik_1(R_1'-R_1)}}{Dk_1^2+Dk_\perp^2-2i|\mu|Q_\bR}
dk_1
\]
\[
=\pi\int_{\bk_\perp}\frac{e^{-|R_1'-R_1|\sqrt{k_\perp^2-2i|\mu|(e^{i{\rm sgn}({\bar Q})\zeta}{\bar Q} +i\eta)/D}}}
{D\sqrt{k_\perp^2-2i|\mu|(e^{i{\rm sgn}({\bar Q})\zeta}{\bar Q} +i\eta)/D}}
\ ,
\eeq
where $\bk_\perp$ is the $\bk$ vector perpendicular to the 1--direction. From the decay of the exponential function 
in this expression we extract a localization length as
\beq
\label{loc_l1}
\xi=\frac{1}{Re\left(\sqrt{k_\perp^2-2i|\mu|(e^{i{\rm sgn}({\bar Q})\zeta}{\bar Q} +i\eta)/D}\right)}
\le c(\eta,|{\bar Q}|,\zeta)\sqrt{D/|\mu|}
\eeq
where the coefficient $c(\eta,|{\bar Q}|,\zeta)$ of the upper bound is of order 1.


\subsection{Higher order terms of the loop expansion}
\label{sect:loops}

The contribution of large loops can be analyzed for the case when we neglect the string term 
$W\varphi W^\dagger\varphi'$ inside the trace of Eq. (\ref{loop_expansion00}), which provides 
the IM without any string contribution as
\beq
J_{loops} =\det({\bf 1}-X\varphi\varphi')\det({\bf 1}+X^\dagger\varphi\varphi')^{-1}
\ .
\eeq
The correlator $\int_\Lambda J_{loops}\varphi_{\bR}\varphi_{\bR'}$ vanishes for $\bR'\ne \bR$, since only
a string formed by $W\varphi W^\dagger\varphi'$ can connect $\varphi_{\bR}$ and $\varphi_{\bR'}$.
On the other hand, in the presence of a string, these loops are obstacles for the string in the full IM $J$ and resemble 
the situation of classical percolation by a geometric restriction of the string formation. 
Reducing the loop distribution to small loops, we have found in Sect. \ref{sect:local} that this leads to localization. 
Therefore, a similar localization effect is anticipated also for a distribution of larger loops, the possibility of
percolating strings for the full distribution of loops cannot be ruled out though.

To understand the interaction between loops and a string, where the latter is created by $W\varphi W^\dagger\varphi'$, 
we consider the spatial off-diagonal elements of the average Hamiltonian $\langle H\rangle$
(i.e., the hopping terms of $\langle H\rangle$). In order to vary the hopping rate we introduce the parameter 
$s$ ($0<s<1$) by the rescaling transformation $\langle H\rangle\to s\langle H\rangle$. For simplicity, we assume that 
$\langle H\rangle$ consists only of nearest-neighbor hopping terms. Then a reduction of $s$ means a reduction of the
hopping probability. Expanding the effective Green's function $h$ of Eq. (\ref{eff_green}) in powers of $s$ gives
\beq
h=\pmatrix{
h_+ & 0 \cr
0 & h_- \cr
}\ , \ \ 
h_\pm=e^{\pm i\phi_\mu}{\bf 1}-\frac{2i\eta}{(\pm\mu-i\eta)^2}s\langle H\rangle +O(s^2)
\]
\[
e^{\pm i\phi_\mu}:=\frac{\pm\mu+i\eta}{\pm\mu-i\eta}
\ .
\eeq
In the PH-symmetric case $\mu=0$
the term linear in the Hermitian matrix $\langle H\rangle$ changes its sign under Hermitian conjugation (i.e., it is
anti-Hermitian). Moreover, we write $X=sX_1+O(s^2)$ and $W=sW_1+O(s^2)$. 
This enables us to extract a scaling factor $s$ in (\ref{loop_expansion00})
and, after rescaling the Grassmann field $\varphi$ by $s$, we obtain
\beq
\log J=
Tr(s[W_1W_1^\dagger]_d\varphi\varphi')
-\sum_{l\ge 1}\frac{1}{l}Tr[(\varphi X_1^\dagger\varphi'+X_1\varphi\varphi'+sW_1\varphi W_1^\dagger\varphi'
\]
\[
+O(s^2))^l]
.
\eeq
Then we treat $s$ as an expansion parameter to separate the terms in $\log J$ according to their scaling
dimension with respect to $s$ as
\beq
\log J=Tr\log({\bf 1}-X_1\varphi\varphi')-Tr\log({\bf 1}+X_1^\dagger\varphi\varphi')
\]
\[
+s\{Tr([W_1W_1^\dagger]_d\varphi\varphi')-Tr(W_1\varphi W_1^\dagger\varphi')\}+O(s^2)
\ .
\eeq
This means that on a scale larger than $1/s$ we can neglect terms of $O(s^2)$. Moreover, since $X_1^\dagger=-X_1$
in the PH-symmetric case, the term without $s$ vanishes, and the leading term on large scales $1/s$ is just
the diffusion propagator linear in $s$. The situation is different when $X_1^\dagger\ne -X_1$ in the PH-symmetry 
broken case. Then the term without $s$ survives and can dominate the diffusion term.

\section{Discussion and conclusion}

The effect of PH-symmetry breaking is characterized by the appearance of small repulsive dimers in a lattice system that
is diffusive at the PH-symmetric point. This means that diffusion, a classical process, is disturbed by a random 
distribution of small obstacles. 
These obstacles are complex (i.e., they have a phase factor) due to the imaginary matrix $\beta$ in Eq. (\ref{loc_matrix}).
This effect can be seen as a quantum effect, since for classical diffusion in a random environment there are only real
obstacles. The dimers are represented by a Gaussian field with a complex correlation matrix in the Grassmann
functional integral, as given in Eqs. (\ref{Gauss_dist}), (\ref{grassmann_int2}).
This reduces the original problem of the IM in Eqs. (\ref{IM0001}) and (\ref{corr00})
to the rather elementary case of diffusion in the presence of random obstacles.
The effect of the latter has been estimated and gives a localization length, whose upper bound is
$\sqrt{D/|\mu|}$ with the diffusion coefficient $D$ and the PH-symmetry breaking parameter $\mu$.
This surprisingly elementary result, which depends only on the ratio of the two model parameters, reflects the 
competition between diffusion and PH-symmetry breaking. The diffusion coefficient, which is determined through
the expression ${\tilde\gamma}_0-{\tilde\gamma}_\bk=Dk^2$, depends on the average Hamiltonian $H_0$,
the scattering rate $\eta$ and $\mu$ according to the expressions in Eq. (\ref{eff_green}).
On the other hand, the localization effect due to PH-symmetry breaking does not agree with the conventional picture 
of a mobility edge somewhere in the band of the random Hamiltonian and the related second 
order phase transition for dimensionality $d>2$~\cite{ab.an.79,wegner79}.
Moreover, our result is different from the self-consistent approach to Anderson localization by Vollhardt and
Wölfle~\cite{PhysRevB.22.4666,wolfle2010}, who found that the diffusion coefficient $D$ vanishes at 
the transition to Anderson localization
but leaves a pole of the effective propagator on the real axis instead of moving it away
into the complex plane. These differences might be related to the fact that we have considered a special class of
Hamiltonians, based on the property (\ref{id0}) and the chiral symmetry. Moreover, the type of PH-symmetry breaking in 
Eq. (\ref{8by8a}) is special. With our results we cannot rule out that there are other types of PH-symmetry breaking which 
lead to the conventional Anderson transition. 

The type of Hamiltonian obeying (\ref{id0}) is known in the form of the Dirac Hamiltonian, which is realized
for low-energy quasi particles in graphene. As mentioned in the Introduction, this material has been the subject of 
intense experimental as well as theoretical research for a number of years. Our result of a finite localization length
in the case of a broken PH symmetry might be useful for the characterization of transport in doped graphene.
For this system
the diffusion coefficient of the two-dimensional Dirac fermions with finite momentum cut-off reads 
$D=\frac{(\hbar v_F)^2}{\eta}\gamma_D$~\cite{zi_13}, where the scalar $\gamma_D$
depends on the momentum cut-off. The Fermi velocity in graphene is typically $v_F\approx10^6$m/sec and
a typical Fermi energy $\mu$ is up to $|\mu|\approx 0.5$eV. Moreover, the typical scattering rate is
$\eta\approx 0.02$eV. Together with the Planck constant
$
\hbar\approx 6.6\cdot 10^{-16}
$eVsec we get for the localization length $\xi\sim \sqrt{D/|\mu|}\approx10^{-8}$m.
One should keep in mind that a finite system size prevents us to distinguish extended states from localized states whose
localization length is larger than the system size.
This could be important for experiments with graphene flakes 
and for numerical simulations of the localization effect away from the Dirac point,  especially for a small PH-symmetry 
breaking parameter $\mu$. In those cases the localization effect should be observable for localization lengths smaller
than the system size.

\vskip0.4cm

\no
{\bf Acknowledgment:}

\no
This research was supported by a grant of the Julian Schwinger Foundation for Physics Research.

\appendix

\section{The invariant measure}
\label{sect:no_sb0}

Since ${\hat S}^k=0$ for $k>2$ we can write for the matrix of the IM
\[
{\hat H}_0+i\epsilon +i\eta e^{2{\hat S}}
={\hat H}_0+i\epsilon +i\eta (1+2{\hat S}+2{\hat S}^2)
={\hat H}_0+i\epsilon -i\eta +2i\eta ({\bf 1} -{\hat S})^{-1}
\]
\[
=({\bf 1} -{\hat S})^{-1}\hat{\cal G}^{-1}({\bf 1}-\hat{\cal G}{\hat S}){\hat G}_0^{-1}
\]
with
\[
{\hat G}_0^{-1}={\hat H}_0+i\epsilon -i\eta
\ ,\ \ 
\hat{\cal G}=
\pmatrix{
h & 0 \cr
0 & h^T \cr
}
\ .
\]
The definition of graded trace is
\[
Trg\pmatrix{
A & B \cr
C & D \cr
}=Tr A-Tr D
\]
and of the graded determinant is
\beq
\label{detg0}
detg\pmatrix{
A & B \cr
C & D \cr
}
=\frac{\det A}{\det D}\det ({\bf 1}-BD^{-1}CA^{-1})
\ ,
\eeq
where the latter implies
\[
detg(\hat{\cal G})=detg({\hat G}_0)=1
\ .
\]
This gives for the IM
\[
J=detg( {\bf 1} -{\hat S})^{-1}detg( {\bf 1}-\hat{\cal G}{\hat S}))
\ .
\]
Moreover, using
\[
{\hat S}=\pmatrix{
0 & {\bar S} \cr
{\bar S}' & 0 \cr
}
,\ 
{\bar S}=\pmatrix{
0 & \varphi_1U \cr
\varphi_2U & 0 \cr
}
=\pmatrix{
\varphi_1U & 0\cr
0 & \varphi_2U \cr
}\sigma_1
,
\]
\[
{\bar S}'=\pmatrix{
0 & \varphi'_2U^\dagger \cr
\varphi'_1U^\dagger & 0 \cr
}
=\sigma_1\pmatrix{
\varphi'_1U^\dagger & 0\cr
0 & \varphi'_2U^\dagger \cr
}
,
\]
we can express the IM via (\ref{detg0}) in terms of determinants as
\beq
\label{Jdet0}
J=\det({\bf 1}+\bar{S}\bar{S}'-h\bar{S}h^T\bar{S}')
\ .
\eeq
With $H_0:=\langle H\rangle$ and
\[
h=\pmatrix{
(H_0+\mu+i\epsilon-i\eta)(H_0+\mu+i\epsilon+i\eta)^{-1} & 0 \cr
0 & (H_0-\mu+i\epsilon-i\eta)(H_0-\mu+i\epsilon+i\eta)^{-1} \cr
}
\]
we have 
$hh^\dagger={\bf 1}+O(\epsilon)$ and
\[
\sigma_1Uh^TU^\dagger\sigma_1=
\pmatrix{
(H_0+\mu-i\epsilon+i\eta)(H_0+\mu-i\epsilon-i\eta)^{-1} & 0 \cr
0 & (H_0-\mu-i\epsilon+i\eta)(H_0-\mu-i\epsilon-i\eta)^{-1} \cr
}=h^\dagger
\ .
\]
This implies for Eq. (\ref{Jdet0})
\beq
\label{Jdet1}
J=\det({\bf 1}+\varphi_1\varphi'_1-h\varphi_1h^\dagger\varphi'_1)
\det({\bf 1}+\varphi_2\varphi'_2-h\varphi_2h^\dagger\varphi'_2)
\ .
\eeq
Finally, we apply the determinant identity to write
\beq
\det({\bf 1}+\varphi_j\varphi'_j-h\varphi_jh^\dagger\varphi'_j)
=\exp\{-\sum_{l\ge 1}\frac{1}{l}Tr[(-\varphi_j\varphi'_j+h\varphi_jh^\dagger\varphi'_j)^l]\}
\ .
\eeq
The properties of the Grassmann variables imply
\[
\sum_{l\ge 1}\frac{1}{l}Tr[(-\varphi_j\varphi'_j+h\varphi_jh^\dagger\varphi'_j)^l]
=Tr(\varphi'_j\varphi_j)-\sum_{l\ge 1}\frac{1}{l}Tr[(\varphi'_jh\varphi_jh^\dagger)^l]
\]
\[
=-\sum_{l\ge 1}\frac{1}{l}Tr[(-\varphi'_j\varphi_j+\varphi'_jh\varphi_jh^\dagger)^l]
\]
such that
\beq
\label{inversion}
\det({\bf 1}+\varphi_j\varphi'_j-h\varphi_jh^\dagger\varphi'_j)
=\det({\bf 1}+\varphi'_j\varphi_j-\varphi'_jh\varphi_jh^\dagger)^{-1}
\ .
\eeq
This is the relation in Eq. (\ref{IM0001}).

\section*{References}
\bibliography{references}
\end{document}